# Effective Strategies for Using Hashtags in Online Communication

Solomiia Fedushko[1]　　　　　　　　　Sofia Kolos[2]

**Abstract**

The features of use of hashtags among students of Lviv were investigated. The list of optimal strategies for using these communicative tools for personal branding is determined. The effective strategy for using hashtags in online communication for the personal and company branding is considered. The results of calculation of effectiveness of hashtags related to # education is calculated. The reports of using hashtag #education in social networks is presented.

## 1.    Introduction

Hashtags are the best means of attracting attention to content. In posts that have at least one hashtag, engagement rates are on average 12.6% higher than in publications without hashtags. Hashtags are an important part of discovery on Instagram, allowing brands to gain exposure to niche groups and specific areas of interest. While they may not drive exponential engagement growth, they give audiences an organic way to discover branded content through the topics and forums that interest them [1]. Proper use of hashtags is a key condition for successful promotion of their content in the most popular social network today which is Instagram, where millions of users daily post new photos, leave comments and likes, follow the photos of other people's lives, promote products and services. Therefore, it's important to use effective hashtag strategies in Instagram for successful personal branding and strategic marketing of firms of any size.

## 2.    Analysis of Researches and Publications

Hashtag was added to the Oxford English Dictionary in June 2014. The term hashtag means a word or phrase preceded by a hash and used to identify messages relating to a specific topic [2]. The hashtag's widespread use began with Twitter but has extended to other social media platforms. In 2007, developer Chris Messina proposed, in a tweet (Fig.1.), that Twitter began grouping topics using the hash symbol. He was inspired by IRC networks that used pound sight to label groups and topics. Twitter initially rejected the idea. But in October 2007 citizen journalists began using the hashtag #SanDiegoFire, at Messina's suggestion, to tweet updates on a series of forest fires in

[1] Department of Social Communication and Information Science, Lviv Polytechnic National University, Ukraine, Lviv, S. Bandery street 12, E-mail: solomiia.s.fedushko@lpnu.ua

[2] Department of Social Communication and Information Science, Lviv Polytechnic National University, Ukraine, Lviv, S. Bandery street 12, E-mail: otikakolosok@gmail.com

San Diego. The practice of hashtaging took off; now users and brands employ hashtags to cover serious political events and entertainment topics alike [3].

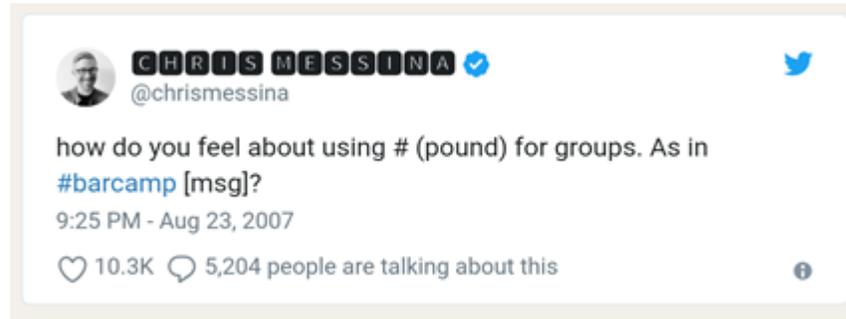

Figure 1 First post with hashtag.

Analysis of sources of information on this problem has shown that the influence of hashtags on the effectiveness of communicative processes in social services is studied by scientists of the different directions. L. Young, T. Sun, M. Ziang analyze the ability of hashtags to unite people into the community. A. Bruns and J. Barbes [4] are studying the performative nature of hashtag addition to the message.

Some scholars carry out quantitative studies of certain hashtag distribution peculiarities in socio-medical networks (MT Bastos, RL Raymondo and R. Traviticky, 2013; Cunya et al., 2011; Ts Ma, A. Sun, and J. Kong, 2012 , D. M. Romero, B. Meder and J. Kylangberg, 2011; O. Zur, A. Rappoport, 2012). J.V. Schurina [5] draws attention to the communicative and gaming potential of the hashtags.

The analysis of the hashtag functions is devoted to J.E. Galimina [6], which highlights, in particular, hashtag-themed markers, hashtag-valued names, hashtags as markers of themselves, predicative-classification and modal function of hashtags.

M. Zappavigena observes that the main functions of the hashtags coincide with the basic functions of the language by M. Holliday [7]: conceptual interpersonal, text. At the same time, the dynamism and constant development of the hashtag's potential require further study of this issue.

## 3. Effective Strategies for Using Hashtags in Online Communication for Personal Branding

The purpose of the study is to compare the results of the questionnaires that characterize the culture of hashtag use in the domestic Signatogram and to determine the optimal strategies for using these communicative means for personal branding.

After conducting the poll within Lviv, authors can say that 87.9% of respondents (Fig.2) know what hashtags are. Also, according to the following statistics, one can see that the social network in which the hashtags are most commonly used is Instagram.

Also, according to the following statistics (Fig.3), one can see that the social network in which the hashtags are most commonly used is Instagram. The second place goes to Facebook.

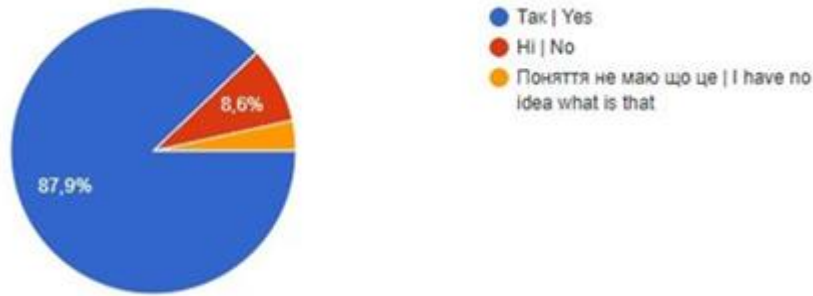

Figure 2 Statistics of the poll in Lviv (first question).

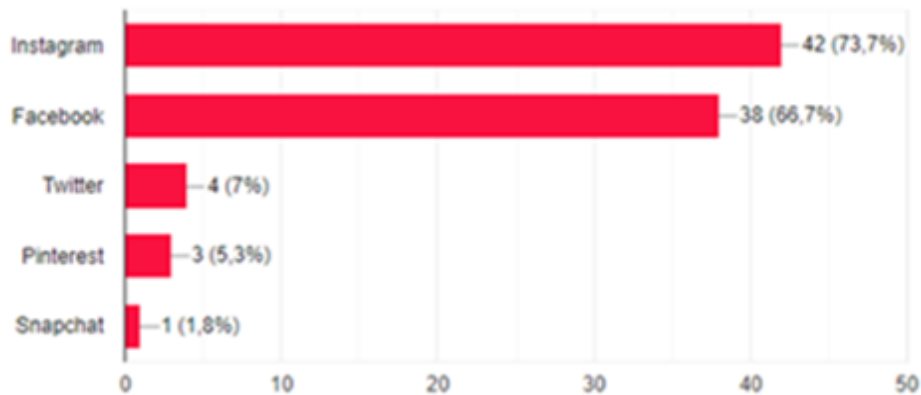

Figure 3 Statistics of the poll in Lviv (second question

Like on Twitter and other social media platforms, Instagram hashtags work by organizing and categorizing photo and video content. Every Instagram post you create can be accompanied by a caption and a few hashtags (up to 30), and these hashtags help in the process of content discovery. In the last year, Instagram has released a few new hashtag features like tracking hashtag engagement with Post Insights, following hashtags, adding clickable hashtags to your Instagram bio and stories.

As we see in statistics, most posts are devoted to photography, and in the second place – traveling. These are the topics of the post, the respondents interviewed most used for personal branding.

For self-branding it is important to take into account the rules for reducing the hashtags because sometimes the name of your brand may be too long and then you need to show creativity, that means you need to convey the basic importance of what you propagate. You can use an abbreviation or an interesting combination of words that relate to your brand, and you can later

Effective Strategies for Using Hashtags in Online Communication

develop a trend by adding this hashtag to all post Fig.4. This can attract new followers and customers and promote your brand on the Internet.

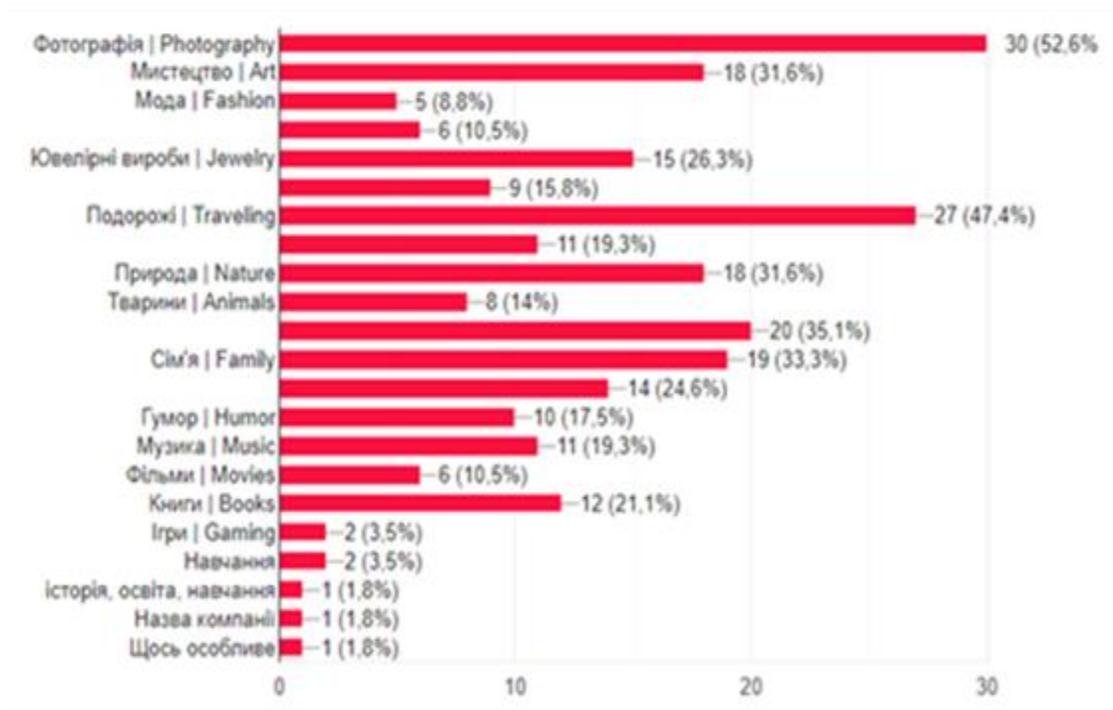

Figure 4 Statistics of the most using topics on Instagram

Based on the analysis of the communicative advantages of the respondents, we recommend that for personal branding on Instagram to take into account the following rules for writing hashtags:

- **Length.** You only need to add a # before a word to make it hashtag. However, because a Tweet is only limited to under 140 characters, the best hashtags are those composed of a single word or a few letters. Twitter experts recommend keeping the keyword under 6 characters. Use only numbers and letters in your keyword. You may use an underscore but do this sparingly for aesthetic reasons.
- **No Spaces**. Hashtags do not support spaces. So if you're using two words, skip the space. For example, hashtags for following the US election are tagged as #USelection, not $US election.
- **No Special Characters**. Hashtags only work with the # sign. Special characters like "!, $, %, ^, &, *, +, ." will not work. But emoji are working as well. Twitter recognizes the pound sign and then converts the hashtag into a clickable link.
- **Don't Start With or Use Only Numbers**. Hashtags like #123 won't work, so don't use only numbers. Similarly, #123yo doesn't work. But numbers are great for recurring events like #conference2012 or #SXSW12.
- **Be careful with slang.** Slang words can mean different things in different countries, so be very mindful about the words you use. Effective hashtags are those that are concise, direct to the point and very relatable across cultures [8].

For personal branding, we consider important effective strategies for using hashtags in Instagram:

- Hashtags should always correspond to the subject of the post.
- To avoid distracting the subscriber from the main - the image itself and not to make posts similar to spam, use 5 hashtags per post.
- Remember that using the most popular hashtags may decrease your personal rating.
- It is important to make individual hashtags and constantly experiment. For example, a good highlight may be a grammar mistake in writing hashtags, if you try.

This is how the domain and the name of the Reglama agency came from the registration and approval of billboards.

## 4. Results

The branding of education and academic companies now-a-days is in demand. The up-to-date researches [9-12] is studied these tasks. The effective strategies for using hashtags in online communication for branding is a probated in Lviv Polytechnic National University official pages in popular social networks.

Reports of competitors of Facebook page of Lviv Polytechnic National University (see Fig. 5), based on sample data of 2 weeks.

The investigation of the general sentiments of hashtag #education (sample data of 2 weeks) in Instagram page of Lviv Polytechnic National University is carried out. Percentage of posts that are positive, neutral and negative, based on the total sample of tracked posts is shown in Fig. 7

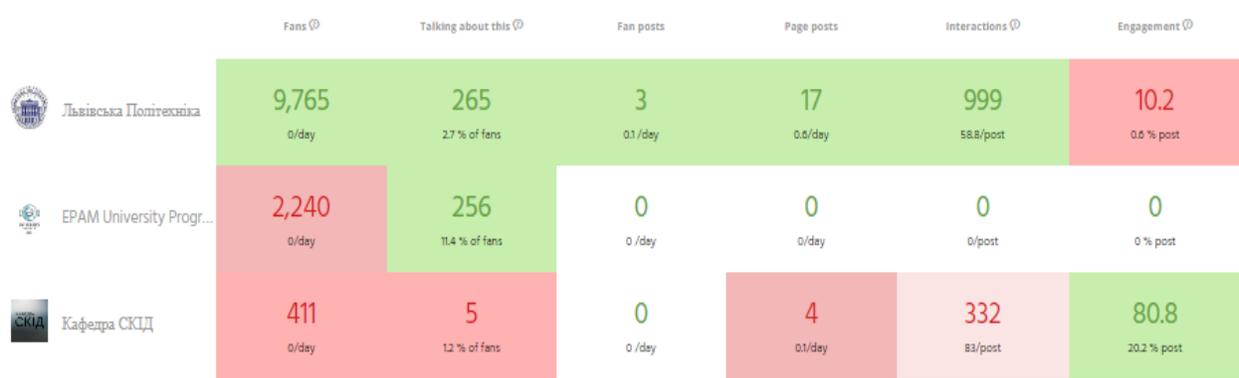

Figure 5 The report of competitors of Facebook page of Lviv Polytechnic National University

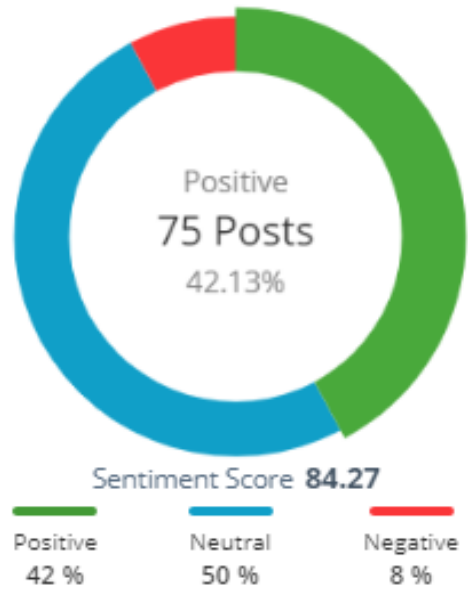

Figure 6 General sentiments of hashtag #education (sample data of 2 weeks) in Instagram page of Lviv Polytechnic National University

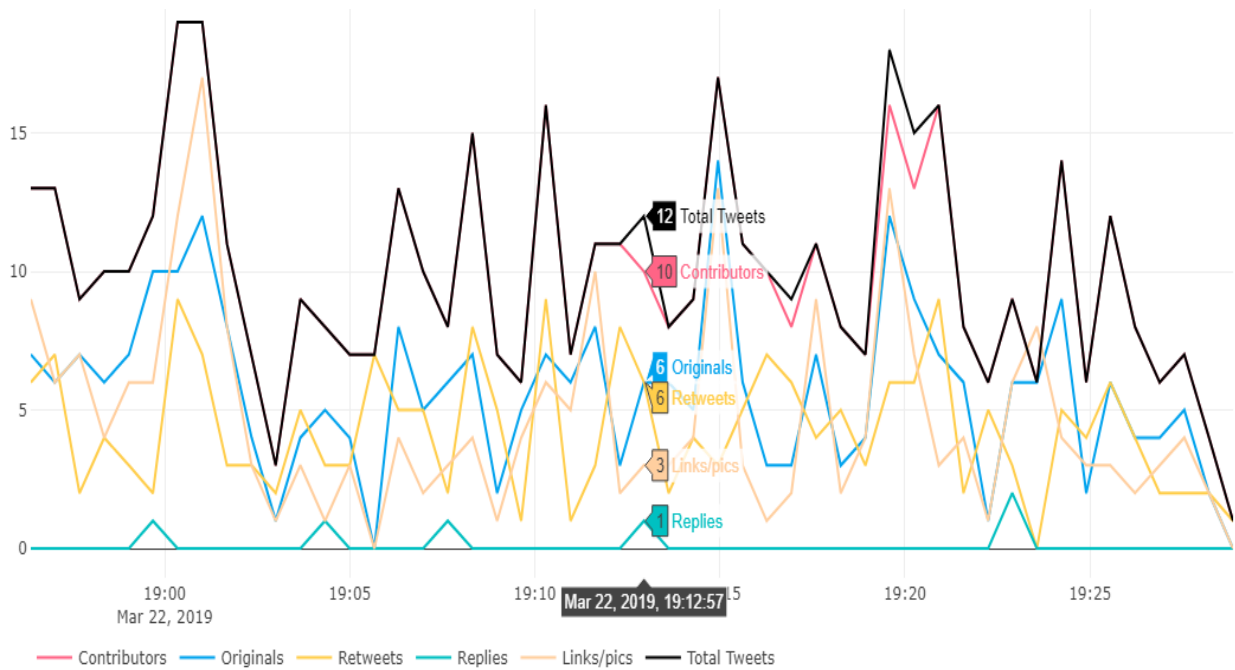

Figure 7 The diagram of using hashtag #education in Tweeter (one day)

Effective Strategies for Using Hashtags in Online Communication

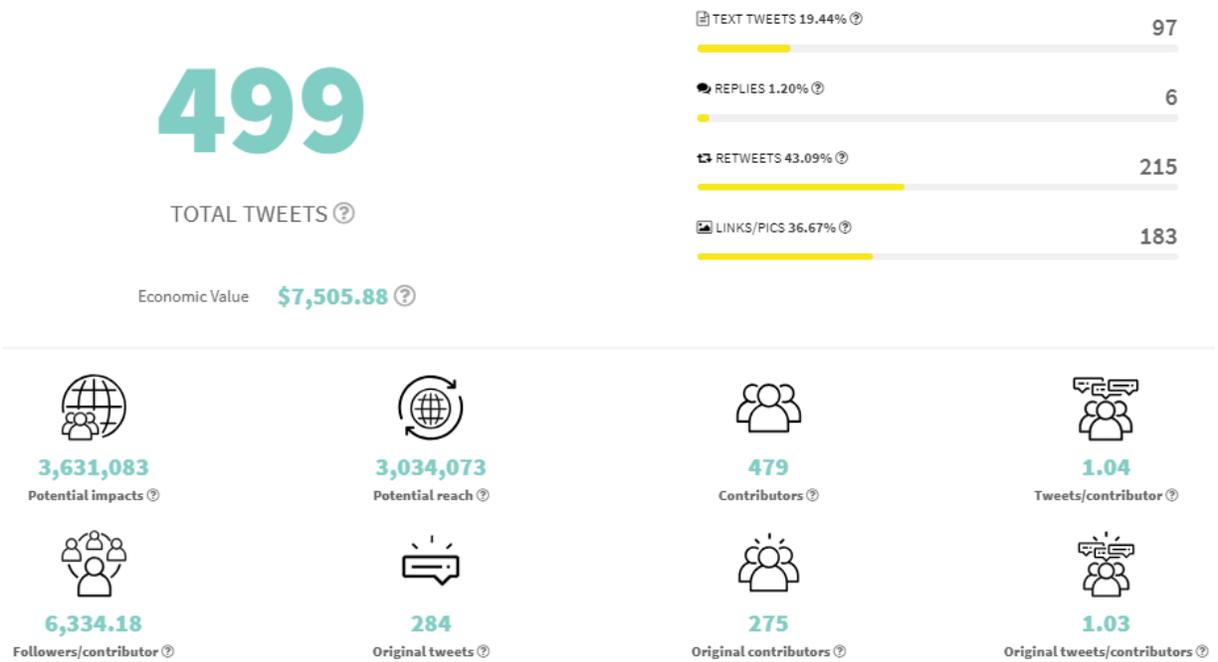

Figure 8 The report of using hashtag #education in Tweeter (for 2 weeks)

This diagram consists the following parameters: contributors, originals, retweets, replies, links/pics and total tweets.

The effectiveness of hashtag is a complex indicator. The effectiveness of hashtag is calculated by the equation (1).

$$Effectiveness^{Hashtag_j}(Post_i) = \begin{Bmatrix} UniqueTweets^{Hashtag_j}, \\ Retweets(Post_i), \\ Exposure^{Hashtag_j}, \\ Relevance^{Hashtag_j}(Post_i) \end{Bmatrix} \quad (1)$$

The results of calculation of effectiveness of hashtags related to #education are presented in Table 1.

Table 1 Effectiveness of hashtags related to #education

| Level of Hashtag Effectiveness | Number | Hashtags |
|---|---|---|
| **High Effectiveness** | 10 | #education, #students, #love, #home, #support, #healthcare, #life, #health, #money, #trading |
| **Average Effectiveness** | 14 | #college, #country, #time, #schools, #apps, #community, #work, #kids, #training, #class, #giving, #career |
| **Low Effectiveness** | 4 | #ubliceducation, #educationsystem, #year, #collegeeducation |

## 5. Conclusion

To this day, building an effective hashtag strategy on Instagram is one of the best ways to get your posts discovered by new audiences on the platform – and depending on how targeted your hashtags are, this can mean more engagement, more followers, and even more customers.